\documentclass{sig-alternate-05-2015}

\usepackage[bookmarks=false]{hyperref}
\usepackage{graphicx}
\usepackage{verbatim}
\usepackage{listings}

\begin{document}

\CopyrightYear{2016}
\setcopyright{acmlicensed}
\conferenceinfo{KDD '16,}{August 13 - 17, 2016, San Francisco, CA, USA}
\isbn{978-1-4503-4232-2/16/08}\acmPrice{\$15.00}
\doi{http://dx.doi.org/10.1145/2939672.2939675}

\clubpenalty=10000 
\widowpenalty = 10000

%
\conferenceinfo{KDD 2016}{8/13-17, San Francisco}

\title{Matrix Computations and Optimization in Apache Spark}

%
%
%
%
%

\numberofauthors{6} 
%
\author{
%
%
\alignauthor
Reza Bosagh Zadeh\titlenote{Corresponding author.}\\
       \affaddr{Stanford and Matroid}\\
       \affaddr{3239 El Camino Real, Ste 310}\\
       \affaddr{Palo Alto, CA 94306}\\
       \email{reza@matroid.com}
\alignauthor
Xiangrui Meng\\
       \affaddr{Databricks}\\
       \affaddr{160 Spear Street, 13th Floor}\\
       \affaddr{San Francisco, CA 94105}\\
       \email{meng@databricks.com}
\alignauthor Alexander Ulanov\\
       \affaddr{HP Labs}\\
       \affaddr{1501 Page Mill Rd}\\
       \affaddr{Palo Alto, CA 94304}\\
       \email{alexander.ulanov@hp.com}    
\and  
\alignauthor Burak Yavuz\\
       \affaddr{Databricks}\\
       \affaddr{160 Spear Street, 13th Floor}\\
       \affaddr{San Francisco, CA 94105}\\
       \email{burak@databricks.com}
\alignauthor Li Pu\\
       \affaddr{Twitter}\\
       \affaddr{1355 Market Street Suite 900.}\\
       \affaddr{San Francisco, CA 94103}\\
       \email{li.pu@outlook.com}
\alignauthor Shivaram Venkataraman\\
       \affaddr{UC Berkeley}\\
       \affaddr{465 Soda Hall}\\
       \affaddr{Berkeley, CA 94720}\\
       \email{shivaram@eecs.berkeley.edu}
\and  
\alignauthor Evan Sparks\\
       \affaddr{UC Berkeley}\\
       \affaddr{465 Soda Hall}\\
       \affaddr{Berkeley, CA 94720}\\
       \email{sparks@cs.berkeley.edu}
\alignauthor Aaron Staple\\
       \affaddr{Databricks}\\
       \affaddr{160 Spear Street, 13th Floor}\\
       \affaddr{San Francisco, CA 94105}\\
       \email{aaron.staple@gmail.com}   
\alignauthor Matei Zaharia\\
       \affaddr{MIT and Databricks}\\
       \affaddr{160 Spear Street, 13th Floor}\\
       \affaddr{San Francisco, CA 94105}\\
       \email{matei@mit.edu}
}

\maketitle
\begin{abstract}
We describe matrix computations available in the cluster programming framework, Apache Spark.
Out of the box, Spark provides abstractions and implementations for
distributed matrices and optimization routines using these matrices. 
When translating single-node algorithms to run on a distributed cluster, we observe that often a simple idea is enough: 
separating matrix operations from vector operations and shipping the matrix operations to be ran on the cluster, 
while keeping vector operations local to the driver. In the case of the Singular Value Decomposition, by taking this idea to an extreme, we are able to
exploit the computational power of a cluster, while running code written decades ago for a single core.
Another example is our Spark port of the popular TFOCS optimization package, originally built for MATLAB, which
allows for solving Linear programs as well as a variety of other convex programs.
We conclude with a comprehensive set of benchmarks for hardware accelerated matrix computations
from the JVM, which is interesting in its own right, as many cluster programming frameworks use the JVM.
The contributions described in this paper are already merged into Apache Spark and
available on Spark installations by default, and commercially supported by a slew of companies which
provide further services.
\end{abstract}

%
%

%
%

%
%


\section{Introduction}

Modern datasets are rapidly growing 
in size and many datasets come in the form of matrices. There is a
pressing need to handle large matrices spread across many machines
 with the same familiar linear algebra tools that are available for single-machine analysis.
Several `next generation' data flow engines that
generalize MapReduce~\cite{mapreduce} have been developed for large-scale data
processing, and building linear algebra functionality on these engines is a
problem of great interest.  In particular, Apache Spark~\cite{spark} has
emerged as a widely used open-source engine.  Spark is a fault-tolerant and
general-purpose cluster computing system providing APIs in Java, Scala, Python,
and R, along with an optimized engine that supports general execution graphs.
 
In this work we present Spark's distributed linear algebra and optimization libraries, 
the largest concerted cross-institution effort to build a distributed linear algebra and optimization library.  
The library
targets large-scale matrices  that benefit from row, column, entry, or block sparsity
 to store and operate on distributed and local matrices. The library, named \textsc{linalg} 
consists of fast and scalable implementations of standard
matrix computations for common linear algebra operations including basic operations such as
multiplication and more advanced ones such as factorizations. It also provides a variety of underlying
primitives such as column and block statistics.
Written in Scala and using native (\verb!C++! and fortran based) linear algebra libraries on each
node, \textsc{linalg} includes Java, Scala, and Python APIs, and is released
as part of the Spark project under the Apache 2.0 license. 

\subsection{Apache Spark}

We restrict our attention to Spark, because it has several
features that are particularly attractive for matrix computations:

\begin{enumerate}
\item The Spark storage abstraction called Resilient Distributed
Datasets (RDDs) is essentially a distributed fault-tolerant vector
on which programmers can perform a subset of operations expected
from a regular local vector.
\item RDDs permit user-defined data partitioning, and
the execution engine can exploit this to co-partition
RDDs and co-schedule tasks to avoid data movement.
\item Spark logs the lineage of operations used to build
an RDD, enabling automatic reconstruction of lost
partitions upon failures. Since the lineage graph is
relatively small even for long-running applications,
this approach incurs negligible runtime overhead,
unlike checkpointing, and can be left on without concern
for performance. Furthermore, Spark supports
optional in-memory distributed replication to reduce
the amount of recomputation on failure.
\item Spark provides a high-level API in Scala that can be
easily extended. This aided in creating a coherent
API for both collections and matrix computations.
\end{enumerate}

There exists a history of using clusters of machines for distributed
linear algebra, for example \cite{scalapack}. These systems are often 
not fault-tolerant to hardware failures and assume 
random access to non-local memory.
In contrast, our library is built on Spark, which is a dataflow system without direct access to non-local memory, 
designed for clusters of \textit{commodity} computers with relatively slow and cheap interconnects, and abundant machines failures.
All of the contributions described in this paper are already merged into Apache Spark and
available on Spark installations by default, and commercially supported by a slew of companies which
provide further services.

\subsection{Challenges and Contributions}

Given that we have access to RDDs in a JVM environment, four key challenges arise 
to building a distributed linear algebra library, each of which we address:

\begin{enumerate}
\item Data representation: how should one partition the entries of
a matrix across machines so that subsequent matrix operations can be implemented as efficiently as possible?
This led us to develop three different distributed matrix representations, each of which has benefits depending
on the sparsity pattern of the data. We have built 
\begin{enumerate}
\item \textsc{CoordinateMatrix} which puts each nonzero into a separate RDD entry. 
\item \textsc{BlockMatrix} which treats the matrix as dense blocks of non-zeros, each block small enough to fit in memory on a single machine.
\item \textsc{RowMatrix} which assumes each row is small enough to fit in memory. There is an option to use a sparse or dense representation for each row.
\end{enumerate}

These matrix types and the design decisions behind them are outlined in Section \ref{sec:representation}.

\item Matrix Computations must be adapted for running on a cluster, as we cannot readily reuse linear algebra algorithms available
for single-machine situations. A key idea that lets us distribute many operations is to separate 
algorithms into portions that require matrix operations versus vector operations. Since matrices
are often quadratically larger than vectors, a reasonable assumption is
that vectors fit in memory on a single machine, while matrices do not. Exploiting this idea, 
we were able to distribute the Singular Value Decomposition via code written decades ago in FORTRAN90, as part
of the ARPACK \cite{arpack} software package. By separating matrix from vector computations, 
and shipping the matrix computations to the cluster while keeping vector operations local to the driver, we were able to
distribute two classes of optimization problems:
\begin{enumerate}
\item Spectral programs: Singular Value Decomposition (SVD) and PCA
\item Convex programs: Gradient Descent, LBFGS, Accelerate Gradient, and other unconstrained optimization methods. 
We  provide a port of the popular TFOCS optimization framework \cite{tfocs}, 
which covers Linear Programs and a variety of other convex objectives 
\end{enumerate}

Separating matrix operations from vector operations helps in the case that vectors fit in memory, but matrices do not. This covers a wide
array of applications, since matrices are often quadratically larger. However, there are some cases for which vectors
do not fit in memory on a single machine. For such cases, we use an RDD for the vector as well, and use BlockMatrix for data storage. 

We give an outline of the most interesting of these computations in Section \ref{sec:computations}.

\item Many distributed computing frameworks such as Spark and Hadoop run on 
the Java Virtual Machine (JVM), which means that achieving hardware-specific acceleration for computation can be 
difficult. We provide a comprehensive survey of tools that allow matrix computations 
to be pushed down to hardware via the Basic Linear Algebra Subprograms (BLAS) interface from the JVM. In addition to
a comprehensive set of benchmarks, we have made all the code for producing the benchmark public to allow
for reproducibility.

In Section \ref{sec:benchmarks} we provide results and pointers to code and benchmarks.

\item Given that there are many cases when distributed matrices and local matrices need to interact (for example
multiplying a distributed matrix by a local one), we also briefly describe the local linear algebra library we built to make this possible, although the focus
of the paper is distributed linear algebra.

\end{enumerate}

\section{Distributed matrix} \label{sec:representation}
Before we can build algorithms to perform distributed matrix computations, we need to lay out the matrix across machines. 
We do this in several ways, all of which use the sparsity pattern to optimize computation and space usage.
A distributed matrix has long-typed row and column indices and double-typed values, stored distributively in one or more RDDs. It is very important to choose the right format to store large and distributed matrices. Converting a distributed matrix to a different format may require a global shuffle, which is quite expensive. Three types of distributed matrices have been implemented so far.

\subsection{RowMatrix and IndexedRowMatrix}

A \textsc{RowMatrix} is a row-oriented distributed matrix without meaningful row indices, backed by an RDD of its rows, where each row is a local vector. Since each row is represented by a local vector, the number of columns is limited by the integer range but it should be much smaller in practice. We assume that the number of columns is not huge for a \textsc{RowMatrix} so that a single local vector can be reasonably communicated to the driver and can also be stored / operated on using a single machine. 

An \textsc{IndexedRowMatrix} is similar to a \textsc{RowMatrix} but with meaningful row indices. It is backed by an RDD of indexed rows, so that each row is represented by its index (long-typed) and a local vector.

\subsection{CoordinateMatrix}

A \textsc{CoordinateMatrix} is a distributed matrix backed by an RDD of its entries. Each entry is a tuple of \textsc{(i: Long, j: Long, value: Double)}, where \textsc{i} is the row index, \textsc{j} is the column index, and \textsc{value} is the entry value. 

A \textsc{CoordinateMatrix} should be used only when both dimensions of the matrix are huge and the matrix is very sparse. A \textsc{CoordinateMatrix} can be created from an \textsc{RDD[MatrixEntry]} instance, where \textsc{MatrixEntry} is a wrapper over \textsc{(Long, Long, Double)}. A \textsc{CoordinateMatrix} can be converted to an \textsc{IndexedRowMatrix} with sparse rows by calling \textsc{toIndexedRowMatrix}.

\subsection{BlockMatrix}

A \textsc{BlockMatrix} is a distributed matrix backed by an RDD of \textsc{MatrixBlock}s, where a \textsc{MatrixBlock} is a tuple of \textsc{((Int, Int), Matrix)}, where the \textsc{(Int, Int)} is the index of the block, and \textsc{Matrix} is the sub-matrix at the given index with size rowsPerBlock $\times$ colsPerBlock. \textsc{BlockMatrix} supports methods such as \textsc{add} and \textsc{multiply} with another \textsc{BlockMatrix}. BlockMatrix also has a helper function validate which can be used to check whether the \textsc{BlockMatrix} is set up properly.

\subsection{Local Vectors and Matrices}

Spark supports local vectors and matrices stored on a single machine, as well as distributed matrices backed by one or more RDDs. Local vectors and local matrices are simple data models that serve as public interfaces. The underlying linear algebra operations are provided by Breeze  and jblas. A local vector has integer-type and 0-based indices and double-typed values, stored on a single machine. Spark supports two types of local vectors: dense and sparse. A dense vector is backed by a double array representing its entry values, while a sparse vector is backed by two parallel arrays: indices and values. For example, a vector \textsc{(1.0, 0.0, 3.0)} can be represented in dense format as \textsc{[1.0, 0.0, 3.0]} or in sparse format as \textsc{(3, [0, 2], [1.0, 3.0])}, where 3 is the size of the vector.

\section{Matrix Computations} \label{sec:computations}

We now move to the most challenging of tasks: rebuilding algorithms from single-core modes of computation to operate
on our distributed matrices in parallel. Here we outline some of the more interesting approaches.

\subsection{Singular Value Decomposition}
The rank $k$ singular value decomposition (SVD) of an $m \times n$ real matrix $A$ is a factorization of the form $A = U \Sigma V^T$, where $U$ is an $m \times k$ unitary matrix, $\Sigma$ is an $k \times k$ diagonal matrix with non-negative real numbers on the diagonal, and $V$ is an $n \times k$ unitary matrix. The diagonal entries $\Sigma$ are known as the singular values. The $k$ columns of $U$ and the $n$ columns of $V$ are called the ``left-singular vectors" and ``right-singular vectors" of $A$, respectively.

Depending on whether the $m \times n$ input matrix $A$ is tall and skinny ($m \gg n$) or square, we use different algorithms
to compute the SVD. In the case that $A$ is roughly square, we use the ARPACK package for computing
eigenvalue decompositions, which can then be used to compute a singular value decomposition via the eigenvalue decomposition of $A^TA$.
In the case that $A$ is tall and skinny, we compute $A^T A$, which is small, and use it locally. In the following two sections with detail the approaches for each of these two cases. Note that the SVD of a wide and short matrix can be recovered from its transpose, which is tall and skinny, and so we do not consider the wide and short case. 

There is a well known connection between the eigen and singular value decompositions of a matrix, that is the two decompositions are the same for positive semidefinite matrices, and $A^TA$ is positive semidefinite with its singular values being squares of the singular values of $A$. So one can recover the SVD of $A$ from the eigenvalue decomposition of $A^TA$. We exploit this relationship in the following two sections.

\subsubsection{Square SVD with ARPACK}

ARPACK is a collection of Fortran77 subroutines designed to solve eigenvalue problems \cite{arpack}. Written many decades
ago and compiled for specific architectures, it is surprising that it can be effectively distributed on a modern commodity cluster.

The package is designed to compute a few eigenvalues and corresponding eigenvectors of a general $n \times n$ matrix $A$. In the local setting, it is appropriate for sparse or structured matrices where structured means that a matrix-vector product requires order $n$ rather than the usual order $n^2$ floating point operations and storage. APRACK is based upon an algorithmic variant of the Arnoldi process called the Implicitly Restarted Arnoldi Method (IRAM). When the matrix $A$ is symmetric it reduces to a variant of the Lanczos process called the Implicitly Restarted Lanczos Method (IRLM). These variants may be viewed as a synthesis of the Arnoldi/Lanczos process with the Implicitly Shifted QR technique. The Arnoldi process only interacts with the matrix via matrix-vector multiplies.

APRACK is designed to compute a few, say $k$ eigenvalues with user specified features such as those of largest real part or largest magnitude. Storage requirements are on the order of $nk$ doubles with no auxiliary storage is required. A set of Schur basis vectors for the desired $k$-dimensional eigen-space is computed which is numerically orthogonal to working precision. The only interaction that ARPACK needs with a matrix is the result of matrix-vector multiplies.

By separating matrix operations from vector operations, we are able to distribute the computations required by ARPACK. An important feature of ARPACK is its ability to allow for arbitrary matrix  formats. This is because it does not operate on the matrix directly, but instead acts on the matrix via prespecified operations, such as matrix-vector multiplies. When a matrix operation is required, ARPACK gives control to the calling program with a request for a matrix-vector multiply. The calling program must then perform the multiply and return the result to ARPACK. By using the distributed-computing utility of Spark, we can distribute the matrix-vector multiplies, and thus exploit the computational resources available in the entire cluster.

Since ARPACK is written in Fortran77, it cannot immediately be used on the Java Virtual Machine. However, through the netlib-java and breeze packages, we use ARPACK on the JVM on the driver node and ship the computations required for matrix-vector multiplies to the cluster. This also means that low-level hardware optimizations can be exploited for any local linear algebraic operations. As with all linear algebraic operations within MLlib, we use hardware acceleration whenever possible. This functionality has been available since Spark 1.1.

We provide experimental results using this idea. A very popular matrix in the recommender systems community is the Netflix Prize Matrix. The matrix has 17,770 rows, 480,189 columns, and 100,480,507 non-zeros. Below we report results on several larger matrices, up to 16x larger.

With the Spark implementation of SVD using ARPACK, calculating wall-clock time with 68 executors and 8GB memory in each, looking for the top 5 singular vectors, we can factorize larger matrices distributed in RAM across a cluster, in a few seconds, with times listed Table \ref{tab:arpack}.

\begin{table*}
\center
\begin{tabular}{|c|c|c|c|}
\hline
\textbf{Matrix size} & \textbf{Number of nonzeros} & \textbf{Time per iteration (s)} & \textbf{Total time (s)} \\ \hline
23,000,000 $\times$ 38,000  & 51,000,000                  & 0.2                             & 10                      \\ \hline
63,000,000 $\times$ 49,000  & 440,000,000                 & 1                               & 50                      \\ \hline
94,000,000 $\times$ 4,000   & 1,600,000,000               & 0.5                             & 50                      \\ \hline
\end{tabular}
\caption{Runtimes for ARPACK Singular Value Decomposition} \label{tab:arpack}
\end{table*}

\subsubsection{Tall and Skinny SVD}

In the case that the input matrix has few enough columns that $A^TA$ can fit in memory on the driver node,
we can avoid shipping the eigen-decomposition to the cluster and avoid the associated communication costs.

First we compute $\Sigma$ and $V$. We do this by computing $A^TA$, which can be done with one all-to-one communication,
details of which are available in \cite{dimsum, disco}. Since $A^TA = V \Sigma^2 V^T$ is of dimension $n \times n$, 
for small $n$ (for example $n=10^4$) we can compute the eigen-decomposition of
$A^TA$ directly and locally on the driver to retrieve $V$ and $\Sigma$.

Once $V$ and $\Sigma$ are computed, we can recover $U$. Since in this case $n$ is small enough 
to fit $n^2$ doubles in memory, then $V$ and $\Sigma$ will also fit in memory on the driver. $U$ however
will not fit in memory on a single node and will need to be distributed, and we still need to compute it. This can be achieved
by computing $U = A V \Sigma^{-1}$ which is derived from $A = U \Sigma V^T$. $\Sigma^{-1}$ is easy to compute since
it is diagonal, and the pseudo-inverse of $V$ is its transpose and also easy to compute. We can distribute the computation of 
$U = A(V \Sigma^{-1} )$ by broadcasting $V \Sigma^{-1} $ to all nodes holding rows of $U$, 
and from there it is embarrassingly parallel to compute $U$.

The method \textsc{computeSVD} on the \textsc{RowMatrix} class takes care of which of the tall and skinny or square versions
to invoke, so the user does not need to make that decision.

\subsection{Spark TFOCS: Templates for First-Order Conic Solvers}

To  allow users of single-machine optimization algorithms to use commodity clusters, we have developed
Spark TFOCS, which is an implementation of the TFOCS convex solver for Apache Spark.

The original Matlab TFOCS library \cite{tfocs} provides building blocks to construct efficient solvers for convex problems. 
Spark TFOCS implements a useful subset of this functionality, in Scala, and is designed to operate on distributed data 
using the Spark. Spark TFOCS includes support for:

\begin{itemize}
\item Convex optimization using Nesterov's accelerated method (Auslender and Teboulle variant)
\item Adaptive step size using backtracking Lipschitz estimation
\item Automatic acceleration restart using the gradient test
\item Linear operator structure optimizations
\item Smoothed Conic Dual (SCD) formulation solver, with continuation support
\item Smoothed linear program solver
\item Multiple data distribution patterns. (Currently support is only implemented for \textsc{RDD[Vector]} row matrices.)
\end{itemize}

The name ``TFOCS" is being used with permission from the original TFOCS developers, who are not involved in the development of this package and hence not responsible for the support. To report issues or download code, please see the project's GitHub page 
\begin{center} \url{https://github.com/databricks/spark-tfocs} \end{center}

\subsubsection{TFOCS}

TFOCS is a state of the art numeric solver; formally, a first order convex solver \cite{tfocs}. This means that it optimizes functions that have a global minimum without additional local minima, and that it operates by evaluating an objective function, and the gradient of that objective function, at a series of probe points. The key optimization algorithm implemented in TFOCS is NesterovÕs accelerated gradient descent method, an extension of the familiar gradient descent algorithm. In traditional gradient descent, optimization is performed by moving ``downhill'' along a function gradient from one probe point to the next, iteration after iteration. The accelerated gradient descent algorithm tracks a linear combination of prior probe points, rather than only the most recent point, using a clever technique that greatly improves asymptotic performance.

TFOCS fine-tunes the accelerated gradient descent algorithm in several ways to ensure good performance in practice, often with minimal configuration. For example TFOCS supports backtracking line search. Using this technique, the optimizer analyzes the rate of change of an objective function and dynamically adjusts the step size when descending along its gradient. As a result, no explicit step size needs to be provided by the user when running TFOCS.

Matlab TFOCS contains an extensive feature set. While the initial version of Spark TFOCS implements only a subset of the many possible features, it contains sufficient functionality to solve several interesting problems.

\subsubsection{Example: LASSO Regression}
A LASSO linear regression problem (otherwise known as L1 regularized least squares regression) 
can be described and solved easily using TFOCS. 
Objective functions are provided to TFOCS in three separate parts, 
which are together referred to as a composite objective function. 
The complete LASSO objective function can be represented as:
$$\frac{1}{2}||Ax-b||_2^2 + \lambda ||x||_1$$
This function is provided to TFOCS in three parts. The first part, the linear component, implements matrix multiplication: $$Ax$$
The next part, the smooth component, implements quadratic loss: $$\frac{1}{2}||\bullet - \space b ||_2^2$$
And the final part, the nonsmooth component, implements L1 regularization: $$\lambda ||x||_1$$
The TFOCS optimizer is specifically implemented to leverage this separation of a composite objective function into component parts. For example, the optimizer may evaluate the (expensive) linear component and cache the result for later use.

Concretely, in Spark TFOCS the above LASSO regression problem can be solved as follows:
\begin{center}
\textsc{TFOCS.optimize(new SmoothQuad(b), new LinopMatrix(A), new ProxL1(lambda), x0)}
\end{center}

Here, \textsc{SmoothQuad} is the quadratic loss smooth component, \textsc{LinopMatrix}
 is the matrix multiplication linear component, and \textsc{ProxL1} is the $L1$ norm nonsmooth component. 
 The \textsc{x0} variable is an initial starting point for gradient descent. 
 Spark TFOCS also provides a helper function for solving LASSO problems, which can be called as follows:
\begin{center}
\textsc{SolverL1RLS.run(A, b, lambda)}
\end{center}

\subsubsection{Example: Linear Programming}

Solving a linear programming problem requires minimizing a linear objective function subject to a set of linear constraints. TFOCS supports solving smoothed linear programs, which include an approximation term that simplifies finding a solution. Smoothed linear programs can be represented as:

\begin{equation*}
\begin{array}{ll@{}ll}
\text{minimize}  & \displaystyle c^Tx  + \frac{1}{2}  ||x- x_0||_2^2  &\\
\text{subject to}& \displaystyle Ax = b\\
                 &    x \geq 0               & 
\end{array}
\end{equation*}
A smoothed linear program can be solved in Spark TFOCS using a helper function as follows:

\begin{center}
\textsc{SolverSLP.run(c, A, b, mu)}
\end{center}

A complete linear program example is presented here:
\begin{center} \url{https://github.com/databricks/spark-tfocs} \end{center}

\subsection{Convex Optimization}

We now focus on Convex optimization via gradient descent for separable objective functions. That is,
objective functions that can be written in the form of $$F(w) = \sum_{i=1}^n F_i(w)$$ where $w$ is a 
$d$-dimensional vector of parameters to be tuned and each $F_i(w)$ represents the loss of the model for the $i$'th training point. 
In the case that $d$ doubles can fit in memory
on the driver, the gradient of $F(w)$ can be computed using the computational resources on the cluster,
and then collected on the driver, where it will also fit in memory. A simple gradient update can be done locally
and then the new guess for $w$ broadcast out to the cluster. This idea is essentially separating the matrix
operations from the vector operations, since the vector of optimization variables is much smaller than the data matrix.

\begin{figure*}
  \centering
    \includegraphics[width=0.45\textwidth]{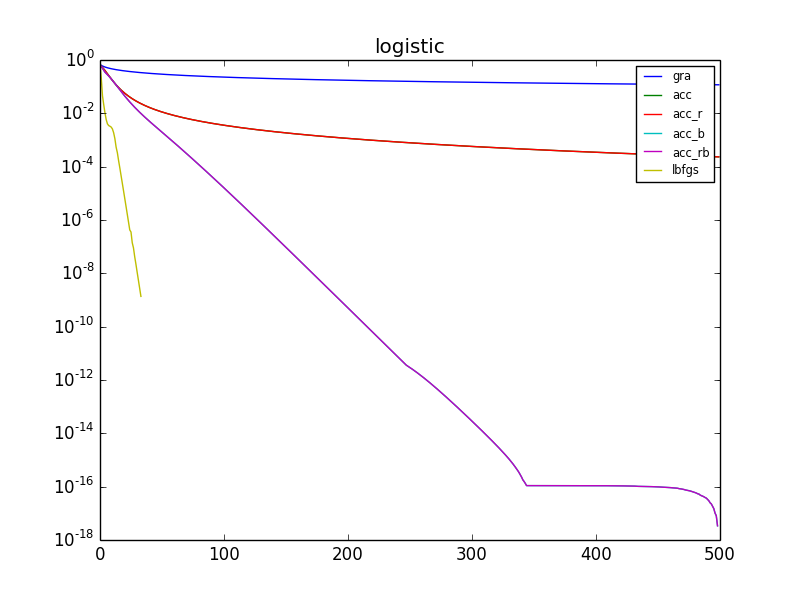}
     \includegraphics[width=0.45\textwidth]{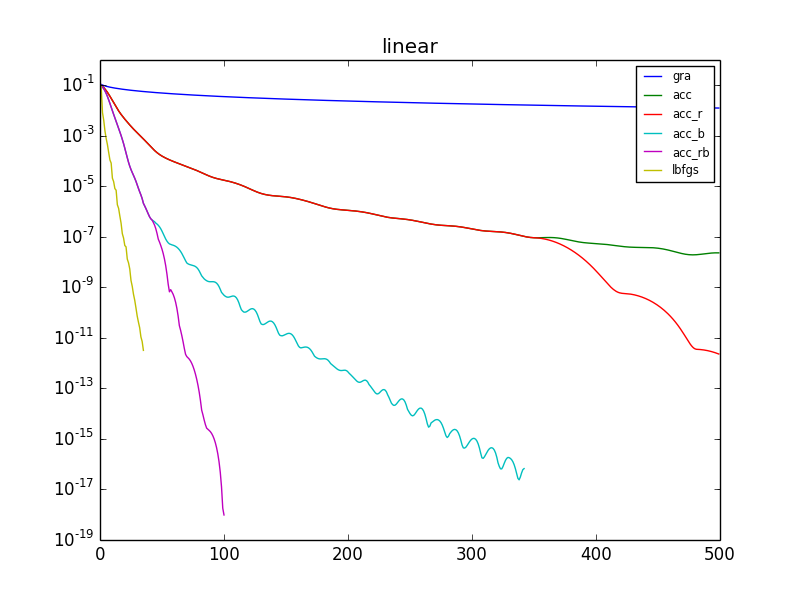}
     \includegraphics[width=0.45\textwidth]{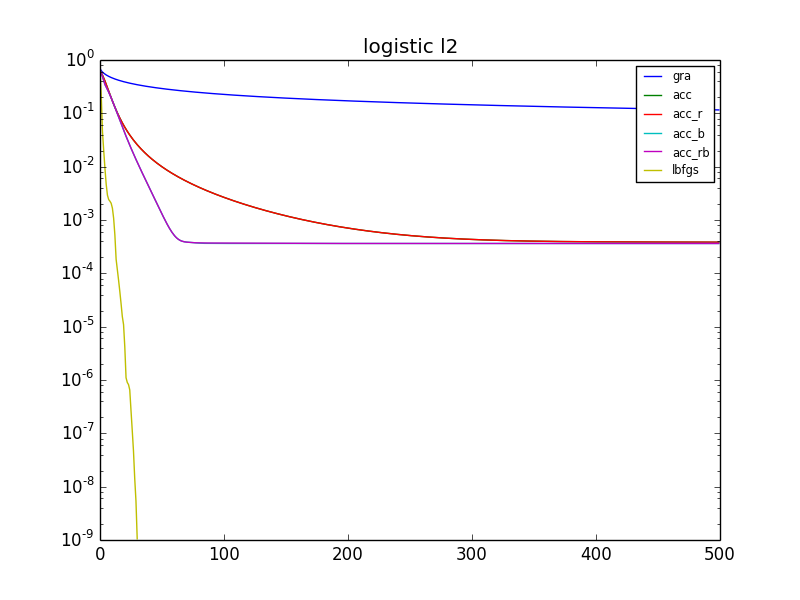}
     \includegraphics[width=0.45\textwidth]{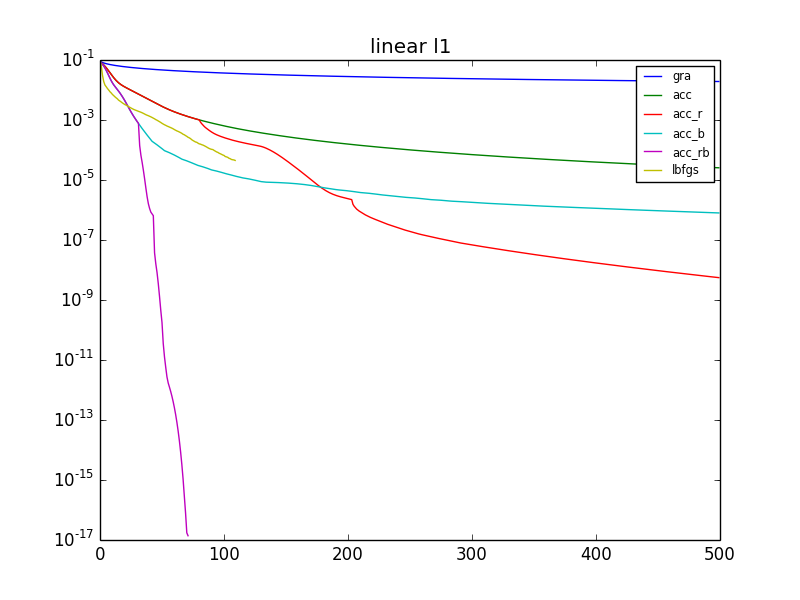}
     \caption{\label{fig:opt} Error per iteration for optimization primitives. From left to right, top to bottom: logistic regression, least squares regression, L2 regularized logistic regression, L1 regularized least squares (LASSO).     }
\end{figure*}

Given that the gradient can be computed using the cluster and then collected on the driver, all computations
on the driver can proceed oblivious to how the gradient was computed. This means in addition to gradient descent,
we can use tradtional single-node implementations of all first-order optimization methods that only use the gradient, such as
accelerated gradient methods, LBFGS, and variants thereof. Indeed, we have LBFGS and accelerated gradient methods
 implemented in this way and available as part of MLlib.  For the first time we provide convergence plots 
 for these optimization primitives available in Spark, listed in Figure \ref{fig:opt}. We have available the following optimization algorithms, with convergence plots in Figure \ref{fig:opt}:

\begin{itemize}
\item gra:  gradient descent implementation \cite{mllib} using full batch
\item acc: accelerated descent as in \cite{tfocs}, without automatic restart \cite{tfocsrestart}
\item acc\_r: accelerated descent, with automatic restart \cite{tfocs}
\item acc\_b: accelerated descent, with backtracking, without automatic restart \cite{tfocs}
\item acc\_rb: accelerated descent, with backtracking, with automatic restart \cite{tfocs}
\item lbfgs: an L-BFGS implementation \cite{lbfgs}
\end{itemize}

In Figure \ref{fig:opt} the $x$ axis shows the number of outer loop iterations of the optimization algorithm. 
Note that for backtracking implementations, the full cost of backtracking is not represented in this outer loop count. 
For non-backtracking implementations, the number of outer loop iterations is the same as the number of spark map reduce jobs. 
The $y$ axis is the log of the difference from best determined optimized value.
The optimization test runs were:
\begin{itemize}
\item linear: A scaled up version of the test data from TFOCS's `test\_LASSO.m' example \cite{tfocs}, with 10000 observations on 1024 features. 512 of the features are actually correlated with result. Unregularized linear regression was used. As expected, the Scala/Spark acceleration implementation was observed to be consistent with the TFOCS implementation on this dataset.
\item linear l1: The same as `linear', but with L1 regularization
\item logistic: Each feature of each observation is generated by summing a feature gaussian specific to the observation?s binary category with a noise gaussian. 10000 observations on 250 features. Unregularized logistic regression was used.
\item logistic l2: Same as `logistic', but using L2 regularization
\end{itemize}

For all runs, all optimization methods were given the same initial step size. We now note some observations. 
First, acceleration consistently converges more quickly than standard gradient descent, given the same initial step size.
Second, automatic restarts are indeed helpful for accelerating convergence.
Third, Backtracking can significantly boost convergence rates in some cases (measured in terms of outer loop iterations), but the full cost of backtracking was not measured in these runs. Finally, LBFGS generally outperformed accelerated gradient descent in these test runs.

\subsection{Other matrix algorithms}
 There are several matrix computations on distributed matrices that use algorithms previously published, so we only cite them here:
 \begin{itemize}
 \item \textsc{RowMatrix} provides QR decomposition as described in \cite{qrpaper}
 \item \textsc{RowMatrix} provides optimized computation of $A^TA$ via DIMSUM \cite{dimsum}
 \item \textsc{BlockMatrix} will provide large linear model parallelism \cite{chen2014large,blockmatrixlinear}
 \end{itemize}

\section{Hardware Acceleration} \label{sec:benchmarks}

\subsection{CPU and GPU acceleration}

To allow full use of hardware-specific linear algebraic operations on a single node, we use the BLAS (Basic Linear Algebra Subroutines) interface with relevant libraries for CPU and GPU acceleration. Native libraries can be used in Scala as follows. First, native libaries must have a C BLAS interface or wrapper. The latter is called through the Java native interface implemented in Netlib-java library and wrapped by the Scala library called Breeze. We consider the following implementations of BLAS like routines:
\begin{itemize}
\item f2jblas - Java implementation of Fortran BLAS
\item OpenBLAS - open source CPU-optimized C implementation of BLAS
\item MKL - proprietary CPU-optimized C and Fortran implementation of BLAS by Intel
\item cuBLAS - proprietary GPU-optimized implementation of BLAS like routines by nVidia. nVidia provides a Fortran BLAS wrapper for cuBLAS called NVBLAS. It can be used in Netlib-java through CBLAS interface.
\end{itemize}

\begin{figure*}
  \centering
    \includegraphics[width=1\textwidth]{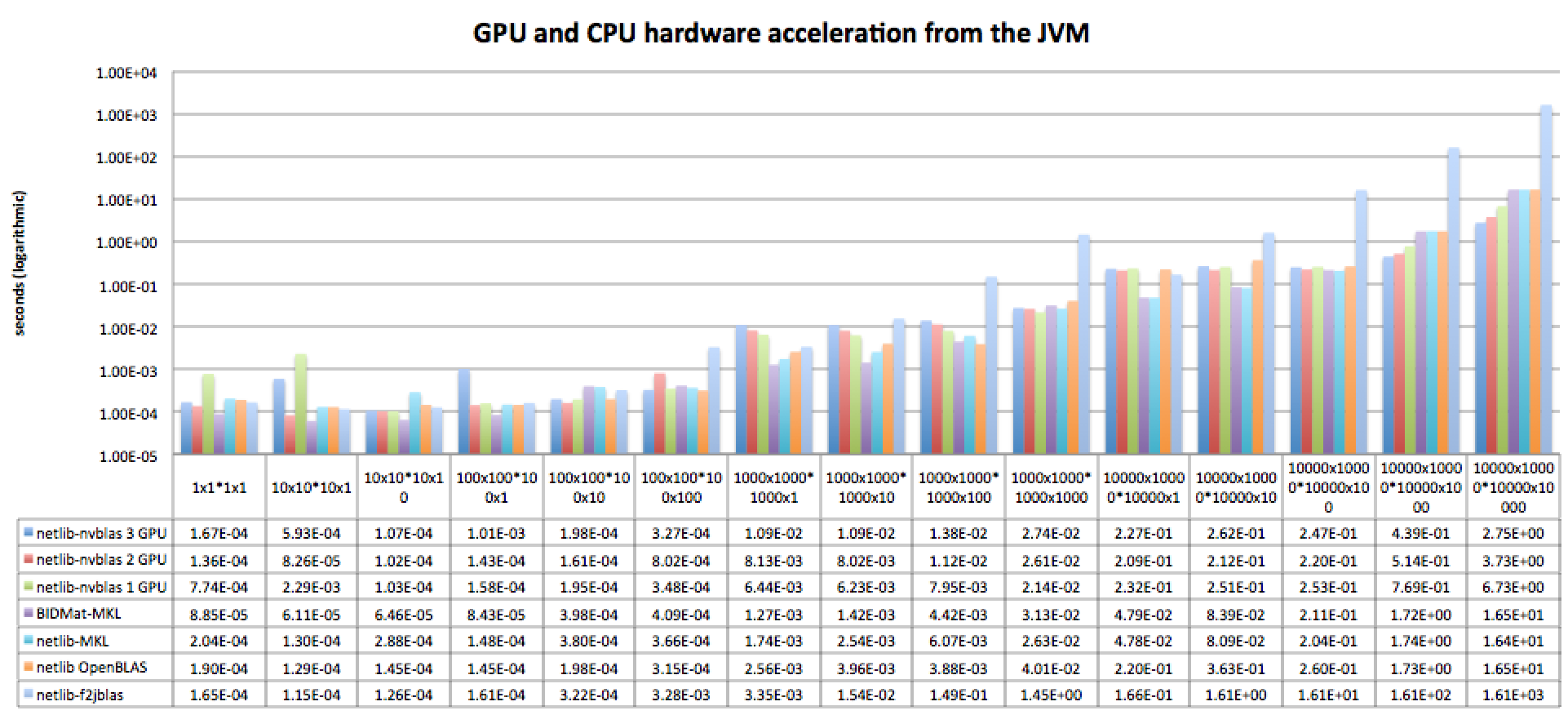}
      \caption{Benchmarks for hardware acceleration from the JVM. Full results and all numbers are available at \url{https://github.com/avulanov/scala-blas}\label{fig:benchmarks}}
\end{figure*}

In addition to the mentioned libraries, we also consider the BIDMat matrix library that can use MKL or cuBLAS.
Our benchmark includes matrix-matrix multiplication routine called GEMM. This operation comes from BLAS Level 3 and can be hardware optimized as opposed to the operations from the lower BLAS levels. We benchmark GEMM with different matrix sizes both for single and double precision. The system used for benchmark is as follows: 
\begin{itemize}
\item CPU - 2x Xeon X5650 @ 2.67GHz (32 GB RAM)
\item GPU - 3x Tesla M2050 3GB, 575MHz, 448 CUDA cores
\item Software - RedHat 6.3, Cuda 7, nVidia driver 346.72, BIDMat 1.0.3
\end{itemize}
The results for the double precision matrices are depicted on Figure \ref{fig:benchmarks}. A full spreadsheet of results and code is available at \url{https://github.com/avulanov/scala-blas}.

Results show that MKL provides similar performance to OpenBLAS except for tall matrices when the latter is slower. Most of the time GPU is less effective due to overhead of copying matrices to/from GPU. However, when multiplying sufficiently large matrices, i.e. starting from 10000$\times$10000 by 10000$\times$1000, the overhead becomes negligible with respect to the computation complexity. At that point GPU is several times more effective than CPU. Interestingly, adding more GPUs speeds up the computation almost linearly for big matrices.

Because it is unreasonable to expect all machines that Spark is run on to have GPUs, we have made OpenBlas the default method of choice for hardware acceleration in Spark's local matrix computations. Note that these performance numbers are useful anytime the JVM is used for Linear Algebra, including Hadoop, Storm, and popular commodity cluster programming frameworks. As an example of BLAS usage in Spark, Neural Networks available in MLlib use the interface heavily, since the forward and backpropagation steps in neural networks are a series of matrix-vector multiplies.

A full spreadsheet of results and code is available at 
\begin{center}
\url{https://github.com/avulanov/scala-blas}
\end{center}

\subsection{Sparse Single-Core Linear Algebra}

The BLAS interface is made specifically for \textit{dense}  linear algebra. There are not many libraries on the JVM that efficiently handle sparse matrix operations, or even provide the option to store a local matrix in sparse format. MLlib provides \textsc{SparseMatrix}, which provides memory efficient storage in Compressed Column Storage (CCS) format. In this format, a row index and a value is stored for each non-zero element in separate arrays. The columns are formed by storing the first and the last indices of the elements for that column in a separate array. 

	MLlib has specialized implementations for performing Sparse Matrix $\times$ Dense Matrix, and Sparse Matrix $\times$ Dense Vector multiplications, where matrices can be optionally transposed.  These implementations outperform libraries such as Breeze, and are competitive against libraries like SciPy, where implementations are backed by C. Benchmarks available at \url{https://github.com/apache/spark/pull/2294}.

\section*{Conclusions}
We described the distributed and local matrix computations available in Apache Spark, a widely distributed cluster programming framework.
By separating matrix operations from vector operations, we are able to distribute a large number of traditional algorithms  meant for single-node usage.
This allowed us to solve Spectral and Convex optimization problems, opening to the door to easy distribution of many machine learning algorithms.
We conclude by providing a comprehensive set of benchmarks on accessing hardware-level optimizations for matrix computations from the JVM.

\section*{Acknowledgments}

We thank all Spark contributors, a list of which can be found at:
\begin{center}
 \url{https://github.com/apache/spark/graphs/contributors}
 \end{center}
 Spark and MLlib are 
cross-institutional efforts, and we thank the Stanford ICME, Berkeley AMPLab, MIT CSAIL, Databricks, Twitter, HP labs, and many other institutions for their support.
We further thank Ion Stoica, Stephen Boyd, Emmanuel Candes, and Steven Diamond for their valuable discussions. 

\bibliographystyle{plain}

\bibliography{linalg}

\appendix

\section*{Appendix A  - BLAS references}

To find information about the implementations used, here we provide links to implementations used in Figure \ref{fig:benchmarks}.

\begin{enumerate}
\item Netlib, Reference BLAS and CBLAS \url{http://www.netlib.org/blas/}
\item Netlib-java \url{https://github.com/fommil/netlib-java}
\item Breeze \url{https://github.com/scalanlp/breeze}
\item BIDMat \url{https://github.com/BIDData/BIDMat/}
\item OpenBLAS \url{https://github.com/xianyi/OpenBLAS}
\item CUDA \url{http://www.nvidia.com/object/cuda_home_new.html}
\item NVBLAS \url{http://docs.nvidia.com/cuda/nvblas}
\end{enumerate}

\end{document}